\documentstyle[11pt,newpasp,twoside]{article}
\reversemarginpar

\begin{document}
\title{Spatially Resolved STIS Spectroscopy of Betelgeuse's Outer Atmosphere}
 \author{A. Lobel, J. Aufdenberg, A. K. Dupree, R. L. Kurucz, R. P. Stefanik,
 \& G. Torres}
\affil{Harvard-Smithsonian Center for Astrophysics, 60 Garden Street, Cambridge 02138 MA, USA}

\begin{abstract}

We present spatially resolved spectra 
observed with the Space Telescope Imaging Spectrograph on the {\em Hubble Space Telescope} of the 
upper chromosphere and dust envelope of $\alpha$ Orionis (M2 Iab). 
In the fall of 2002 a set of five high-resolution near-UV 
spectra was obtained by scanning at intensity peak-up position and 
four off-limb target positions up to one arcsecond, using a small aperture 
(200 by 63 mas), to investigate the thermal conditions and flow 
dynamics in the outer atmosphere of this important nearby cool supergiant star. 

Based on Mg~{\sc ii} $h$ \& $k$, Fe~{\sc ii} $\lambda$2716, C~{\sc ii}
$\lambda$2327, and Al~{\sc ii} ] $\lambda$2669
emission lines we provide the first evidence for the presence of warm chromospheric plasma 
at least 1 arcsecond away from the star at $\sim$40 $\rm R_{*}$ (1 $\rm R_{*}$$\simeq$700 $\rm R_{\odot}$).
The STIS spectra reveal that Betelgeuse's upper chromosphere extends far beyond the circumstellar 
H$\alpha$ envelope of $\sim$5 $\rm R_{*}$, determined from previous
ground-based imaging (Hebden et al. 1987). 

The flux in the broad and self-absorbed resonance lines of Mg~{\sc ii} 
decreases by a factor of $\sim$700 compared to the flux at chromospheric disk center. 
We observe strong asymmetry changes in the Mg~{\sc ii} $h$ and Si~{\sc i}
resonance line profiles 
when scanning off-limb, signaling the outward acceleration of gas outflow in the upper chromosphere. 

From the radial intensity distributions of Fe~{\sc i} and Fe~{\sc ii} emission lines
we determine the radial non-LTE iron ionization balance. We compute that the local kinetic gas 
temperatures of the warm chromospheric gas component in the outer atmosphere 
exceed 2600~K, when assuming local gas densities of the cool gas 
component we determine from radiative transfer models that 
fit the 9.7 $\mu$m silicate dust emission feature. The spatially resolved  STIS spectra directly 
demonstrate that warm chromospheric plasma co-exisists with cool gas in Betelgeuse's 
circumstellar dust envelope. 
\end{abstract}
\begin{figure}
\plotfiddle{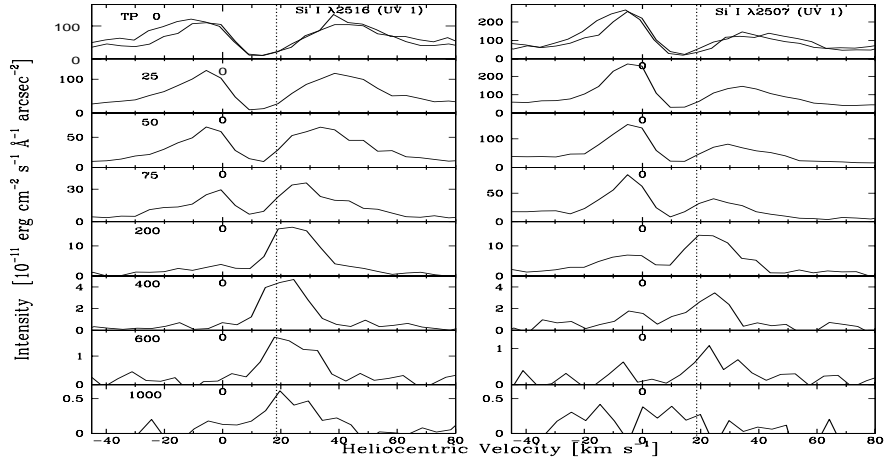}{5cm}{-90}{45}{30}{-165}{170}
\caption{Spatiallly resolved STIS spectra of Si~{\sc i} $\lambda$2516 and
Si~{\sc i} $\lambda$2507 resonance emission lines  observed in Betelgeuse out to 1 arcsecond.   }
\end{figure}
\vspace*{-0.6cm}
\section{STIS Observations}
STIS spectra of the red supergiant $\alpha$ Ori have been observed 
for GO 9369 in HST Cycle 11; {\em A direct Test for Dust-driven Wind Physics}. 
This program investigates the detailed acceleration 
mechanisms of wind outflow in the outer atmospheres (chromosphere and dust
envelope) of cool stars. Using the exceptional capabilities of HST-STIS 
we observe the near-UV spectrum with $\lambda$/$\Delta$$\lambda$$\simeq$33,000 
between 2275 \AA\, and 3180 \AA\,
with spatially resolved scans across the chromospheric disk at 0, 200, 400, 600, \&
1000 mas (Visit 1), at 0 \& 2000 mas (Visit 2), and at 0 \& 3000 mas (Visit 3). 
We presently discuss the spectra observed in fall 2002 of Visit 1.
The spectra of Visits 2 \& 3 of spring 2003 will be presented elsewhere.
Exposure times range from 500 s at 200 mas to 7200 s at 1$\arcsec$, 
yielding good S/N$\geq$20. The spectra are calibrated 
with CALSTIS v2.12 using the most recently updated calibration reference files. 
Wavelength calibration accuracies are better than $\sim$1 detector pixel
or 1.3 $\rm km\,s^{-1}$.
\vspace*{-0.4cm}
\section{Si~{\sc i} $\lambda$2516 line profile changes} 
In previous work we modeled the detailed shape of the Si {\sc i} $\lambda$2516 resonance emission line
(Lobel \& Dupree 2001). The line has previously been observed by scanning 
over the inner chromosphere at 0, 25, 50, and 75 mas, using a slit size of 100
$\times$ 30 mas ({\em Fig. 1, panel left}).
The figure at {\tt cfa-www.harvard.edu/cfa/ep/ \\
pressrel/alobelimg.html} shows the line profiles of Fe~{\sc ii} $\lambda$2869 
for the respective slit positions compared to 
the near-UV continuum flux observed with HST-FOC. The double-peaked
line profiles are observed across the inner chromosphere. The central 
(self-) absorption core results from scattering opacity in the chromosphere. The asymmetry
of the emission component intensities probes the chromospheric flow dynamics in our 
line of sight. The spectra of GO 9369 are observed across the outer chromosphere using 
a slitsize of 200 $\times$ 63 mas ({\em Fig. 1}). The profiles beyond 200 mas {\em appear} 
red-shifted with a rather weak short-wavelength emission component. It signals substantial 
wind outflow opacity in the upper chromoshere, which fastly accelerates beyond
a radius of $\sim$8 $\rm R_{*}$.  
\begin{figure}
\plotfiddle{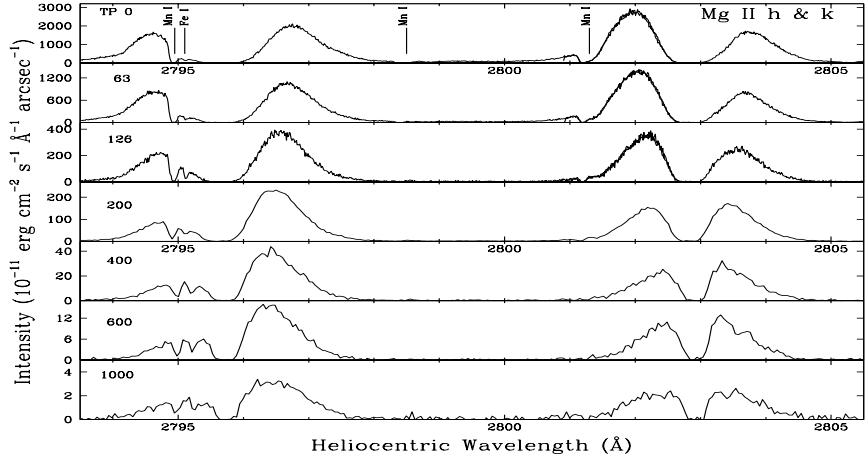}{5cm}{0}{45}{32}{-150}{-25}
\caption{Spatially resolved STIS spectra of Mg~{\sc ii} $h$ \& $k$ emission
  lines observed in the chromosphere of Betelgeuse out to 1 arcsecond.}
\end{figure}
\vspace*{-0.4cm}
\section{Mg~{\sc ii} $k$ \& $h$ line profile changes}
Figure 2 shows the detailed profiles of the Mg~{\sc ii} $h$ \& $k$ 
lines observed up to 1000 mas. The emission line intensities decrease by a factor 
of $\sim$700 from chromospheric disk center (TP 0) to 1$\arcsec$.  
These optically thick chromospheric lines show remarkable changes of their 
detailed shapes when scanning off-limb. The full width across both emission components 
at half intensity maximum decreases by $\sim$20\%, while the broad and saturated central absorption 
core narrows by more than 50\%. Beyond 600 mas the central core assumes a constant width which
results from absorption contributions by the local interstellar medium 
($d_{*}$$\simeq$132 pc). We observe a strong increase 
of the (relative) intensity of the long-wavelength 
emission component in both lines beyond 200 mas.
It signals fast wind acceleration beyond 
this radius. Note that the short-wavelength emission components of the $k$ and $h$ lines are
blended with chomospheric Mn~{\sc i} lines (decreasing the $k$- and increasing the $h$-component), 
but that become much weaker in the outer chromosphere. 
\vspace*{-0.5cm}
\section{Wind Acceleration in the Upper Chromosphere}
Figure 1 compares the profiles of the Si~{\sc i} 
$\lambda$2516 and $\lambda$2507 resonance lines (vertical dotted lines are drawn at stellar rest 
velocity). Both lines share a common upper energy level and their intensities are influenced by pumping 
through a fluoresced Fe~{\sc ii} line. The self-absorption cores of the Si~{\sc i} lines are 
therefore observed far out, into 
the upper chromosphere. The shape of these unsaturated emission lines is strongly opacity sensitive
to the local chromospheric velocity field. As for the Mg~{\sc ii} lines, the outward decreasing intensity of the 
short-wavelength emission component signals fast acceleration of chromospheric outflow in the 
upper chromosphere. We also observe this decrease for the resonance line of Mg~{\sc i} 
$\lambda$2852 (not shown). Our previous radiative transfer modeling work based on Si~{\sc i} revealed 
that $\alpha$ Ori's inner chromosphere oscillates non-radially, with simultaneous up- and downflows 
in Sept. 1998. Radiative transfer modeling to determine the detailed wind structure in the 
outer chromosphere is underway.    
\begin{figure}
\hspace*{1.0cm} 
\plotfiddle{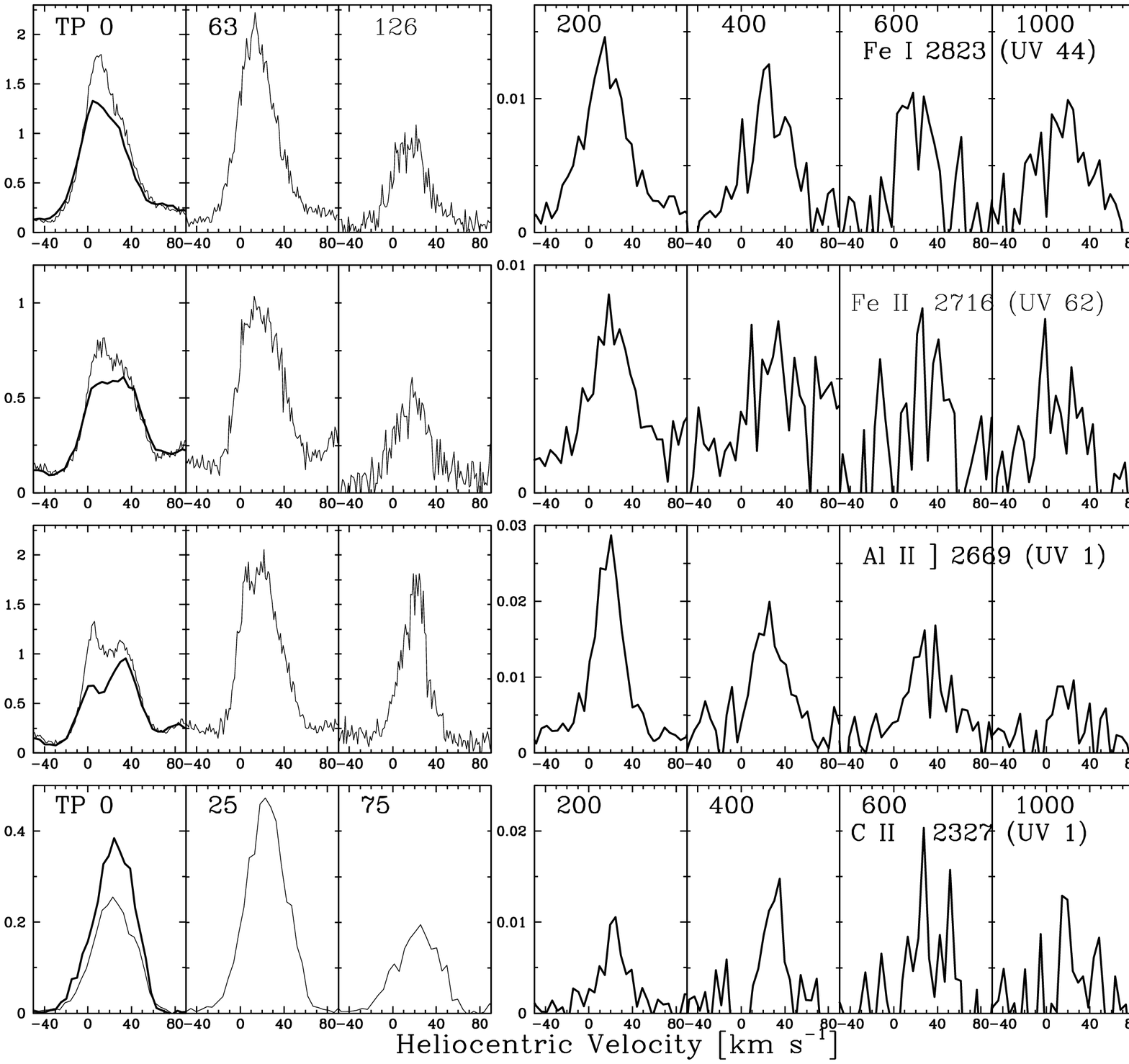}{4cm}{0}{27}{30}{-180}{-40} 
\end{figure}  
\begin{figure}
\hspace*{1.0cm}
\plotfiddle{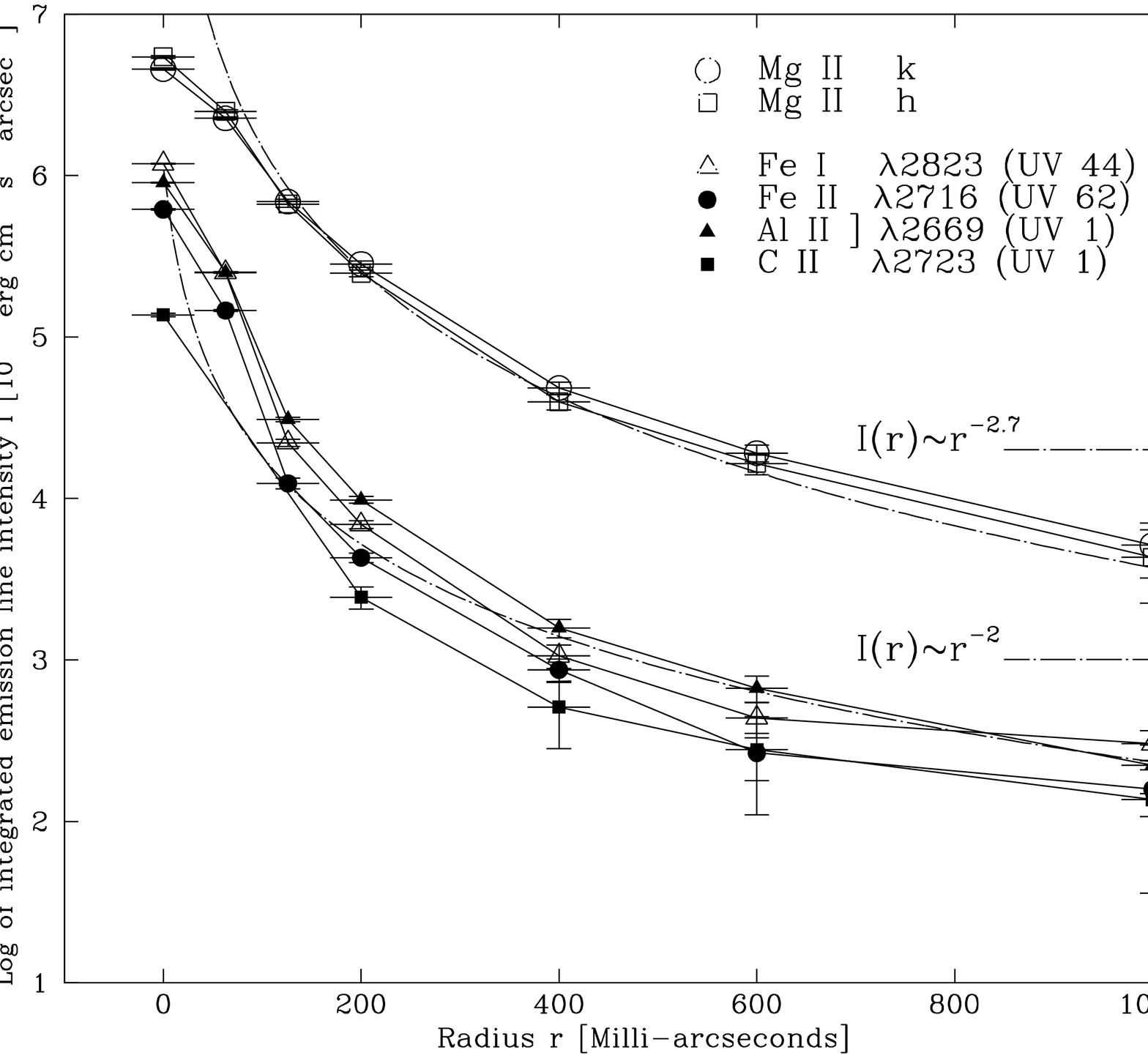}{0cm}{0}{26}{32}{10}{-25} 
\vspace*{0.2cm}
\caption[]{{\em panel left:} Spatially resolved chromospheric emission lines
  of Fe~{\sc i}, Fe~{\sc ii}, Al~{\sc ii} ], and C~{\sc ii}. Figure 4. {\em
  panel right}: Radial chromospheric intensity distribution of these lines
  compared to Mg~{\sc ii} (Fig.~2).}
\end{figure}  
\vspace*{-0.5cm} 
\section{Ion lines in the Upper Chromosphere}
We also observe ion lines of Fe~{\sc ii}, Al~{\sc ii}, and C~{\sc ii} out to 1$\arcsec$ in the 
upper chromosphere. Figure  3 shows (scaled) emission lines of Fe~{\sc ii} 
$\lambda$2716 (UV 62), Al~{\sc ii} ] $\lambda$2669 (UV 1), and C~{\sc ii} $\lambda$2327 (UV 1).
The Fe~{\sc i} $\lambda$2823 (UV 44) line is also shown for comparison ({\em top panels}).
The lines at the inner chromosphere are observed in April 1998 ({\em thin drawn lines}) with 
R$\sim$114,000 at TPs 0, 63, and 126 mas, while the lines of the outer chromosphere are observed with medium resolution 
in fall 2002 ({\em boldly drawn lines}). Both raster scans are however observed with the same slitsize of 200 $\times$ 63 mas
so that the line intensity changes can be compared. For this purpose we select unblended lines
without central self-absorption cores that become sufficiently optically thin in the outer
chromosphere, and that are significantly observed against the local background noise level.
\vspace*{-0.5cm}
\section{Radial Intensity Distribution of Ion Lines}
We wavelength integrate the selected chromospheric emission lines and the Mg~{\sc ii}
lines beyond the line wings. Their radial intensity distributions I(r) are
compared in Fig. 4.   
The intensity errorbars are derived from the STIS pipeline flux calibration errors, while the radius
errorbars are derived from the projected slitwidth. We observe that the I(r) of optically thin
emission from neutral and ion lines are very similar across the chromosphere. Neutral 
emission lines are generally observed farther out with larger S/N compared to the ion lines, but their
I(r) do not differ significantly within the errors. We find a best fit for I(r)$\simeq$const~$\times$~$\rm r^{-2}$.
The I(r) of the optically thick and self-absorbed Mg~{\sc ii} lines differs significantly with 
I(r)$\simeq$const $\times$ $\rm r^{-2.7}$. The steeper intensity distribution signals important radiative 
transfer effects for the shapes of the stronger Mg~{\sc ii} lines (see Sect. 3).  
\vspace*{-0.5cm}
\section{Radial Non-LTE Iron Ionization Balance}
In the upper panel of Fig. 5 we compute the iron ionization fraction
from the I(r) of the Fe~{\sc i} and Fe~{\sc ii} lines. The intersection of the curves
(at dots) provides the excitation temperature corresponding to the observed line intensity
ratios for spontaneous emission. We compute iron ionization fractions between 99.3\% and 99.7\%
for kinetic gas temperatures between 2600 K and 5800 K, using local gas densities 
$10^{-17}$$\leq$$\rho$$\leq$$10^{-15}$ $\rm gr\,cm^{-3}$ ({\em lower panel}). 
This temperature range corresponds to partial NLTE iron ionization due to a 
diluted radiation field with $T_{\rm rad}$$\simeq$3000 K ({\em full drawn lines}), typical for the outer chromosphere.  
The graphs are computed with volume filling factors $\phi$ for warm plasma of
5\% ({\em solid dots}) and 30\% ({\em solid triangles}).
Hydrogen is almost neutral for these conditions in the upper chromosphere.
In Fig. 6 we model the circumstellar dust envelope (CDE) 
with radiative transfer in spherical geometry using {\sc DUSTY}. 
A best fit to the IRAS silicate dust emission feature at 9.7 
$\mu$m yields a dust condensation radius of $R_{c}$$\simeq$573 mas, where 
$\rho_{\rm gas}$$\sim$5 $\times$ $10^{-16}$ $\rm gr\,cm^{-3}$ for the cool ambient gas, with temperatures
below $T_{\rm dust}$$\leq$600 K ({\em lower panels}). The upper panel shows the 
temperature structure for warm chromospheric plasma computed at this $\rho_{\rm gas}$ with 
$\phi$= 1\% and 100\% ({\em bold solid lines}). The inner chromosphere is computed 
with radiative transfer fits to H$\alpha$ (Lobel \& Dupree 2000). 
We find that the temperature of 
warm chromospheric plasma cannot decrease to below 2600 K in the CDE. 
Hence warm chromospheric plasma must co-exisit with cool gas of $T$$\leq$600 K beyond 600 mas.
\setcounter{figure}{4}
\begin{figure}
\hspace*{1.0cm} 
\plotfiddle{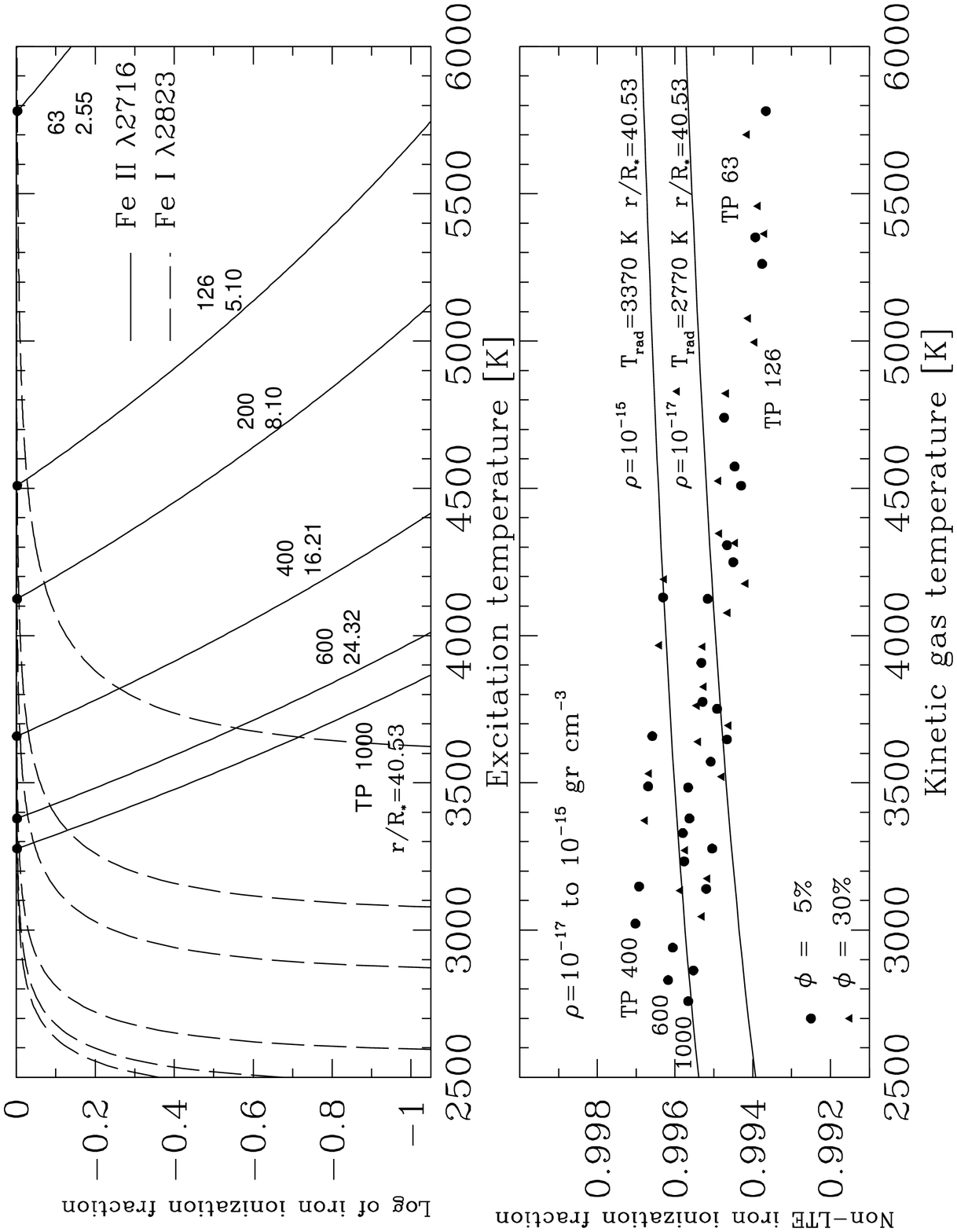}{4cm}{-90}{25}{31}{-200}{142} 
\end{figure}  
\begin{figure}
\hspace*{1.0cm}
\plotfiddle{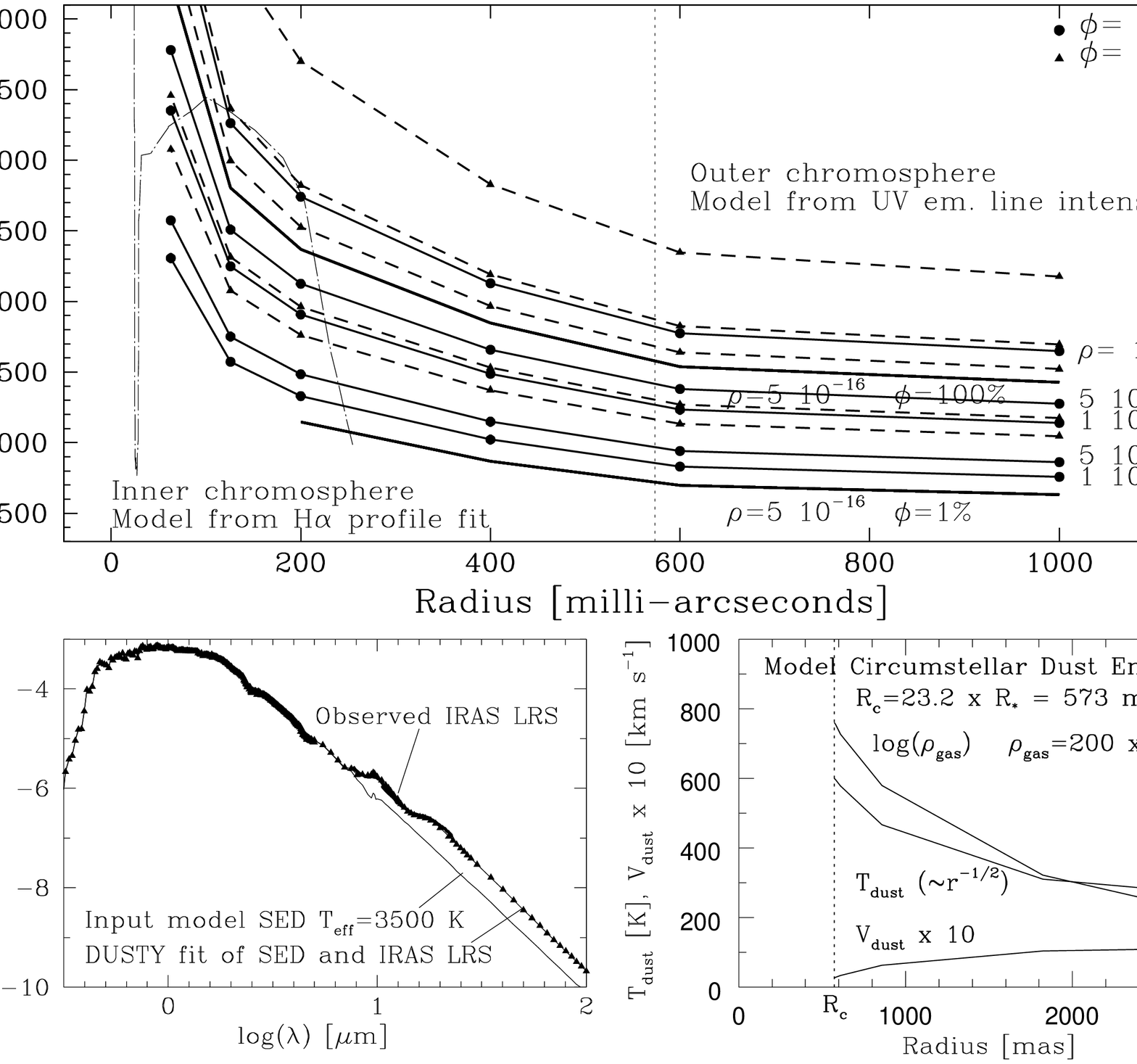}{0cm}{0}{25}{30}{15}{-17} 
\vspace*{0.2cm}
\caption[]{{\em panels left:} Non-LTE iron ionization fraction with kinetic gas
  temperature determined from I(r). Figure 6. {\em panels right:} Temperature
  structure of the chromosphere computed for different densities.}   
\end{figure}  
\vspace*{-0.5cm}
\section{Conclusions}
Spatially resolved spectra of Betelgeuse's outer chromosphere 
signal that it accelerates outwards, based on asymmetries 
observed in Mg~{\sc ii} $h$ \& $k$, and other emission lines.
They also reveal warm chromospheric plasma out to 40 $\rm R_{*}$. 
The chromospheric gas must co-exist with cool gas of the circumstellar dust envelope.

\acknowledgments
This reseach is based on data obtained with the NASA/ ESA Hubble Space
Telescope, collected at the STScI, operated by AURA Inc., under contract
NAS5-26555. Financial support is provided by STScI grant HST-GO-09369.01 to 
the Smithsonian Astrophysical Observatory.

\end{document}